%% file: jstqe_review_final_copyright.tex
\documentclass[journal]{IEEEtran} 
\usepackage{hyperref}
\usepackage{latexsym}
\usepackage{mathrsfs,amssymb,textcomp}

\usepackage{cite}
\ifCLASSINFOpdf
   \usepackage[pdftex]{graphicx}
\else
\fi
\usepackage[cmex10]{amsmath}
\usepackage{array}

\begin{document}
\title{Simulation of nanostructure-based high-efficiency solar cells: challenges, existing
approaches and future directions}

\author{Urs~Aeberhard% <-this % stops a space
\thanks{Urs Aeberhard is with the Institute of Energy and Climate 5: Photovoltaics, Research Centre J\"ulich,
52425 J\"ulich, Germany, e-mail: u.aeberhard@fz-juelich.de.}% <-this % stops a space
}


\markboth{Journal of Selected Topics in Quantum Electronics,~Vol.~, No.~, ~}%
{Shell \MakeLowercase{\textit{et al.}}: Bare Demo of IEEEtran.cls for Journals}


\IEEEpubid{\copyright~2013 IEEE}

\IEEEspecialpapernotice{(Invited Paper)}

\maketitle

\begin{abstract}
Many advanced concepts for high-efficiency photovoltaic devices exploit the peculiar optoelectronic properties of
semiconductor nanostructures such as quantum wells, wires and dots. While the optics of such devices is only modestly
affected due to the small size of the structures, the optical transitions and electronic transport can strongly deviate
from the simple bulk picture known from conventional solar cell devices. This review article discusses the challenges
for an adequate theoretical description of the photovoltaic device operation arising from the introduction of
nanostructure absorber and/or conductor components and gives an overview of existing device simulation approaches.
\end{abstract}

\begin{IEEEkeywords}
solar energy, photovoltaic effect, nanostructures, simulation
\end{IEEEkeywords}

% \fi
\IEEEpeerreviewmaketitle

\section{Introduction}
\IEEEPARstart{T}{he} past decade has witnessed a strong increase in the demand of clean electricity
production based on renewable energy sources and an associated exponential growth of installed
photovoltaic (PV) power capacity. The largest share of the actual PV electricity supply is
still based on crystalline silicon wafer solar cells (SC). However,  this technology, though mature
and highly developed, suffers from several fundamental limitations.
First of all, the energy conversion efficiency of this type of single junction SC device is
limited by the fundamental laws of thermodynamics \cite{shockley:61} to below 40\% even at full
concentration. Due to the low absorption originating in the indirect nature of optical transitions
in crystalline silicon, a large amount of semiconductor material of high quality is required to approach the limiting efficiency.
Furthermore, with the successful reduction of manufacturing costs via optimization of the individual production
steps at industrial scale, efficiency becomes a central issue. This demand for higher PV
energy conversion efficiencies has in the past decade led to the emergence of a whole new generation
of SC concepts \cite{marti:04,green:06}, which all aim at exceeding the single junction
efficiency limit through the reduction of fundamental losses. Well-known representatives are the
concepts based on enhanced spectrum utilization and reduced thermalization losses via the use of
multiple junctions\cite{yamaguchi:05}, intermediate bands \cite{luque:97}, multiple exciton generation \cite{nozik:05}
or hot carrier effects \cite{ross:82}.

While these concepts differ widely in the physical mechanisms exploited, what they have in common is
that they are largely based on artificially engineered materials with designed optoelectronic
properties, like semiconductor nanostructures (NS) such as quantum wells, wires and dots, offering
\mbox{size-,} geometry- and composition-tunable characteristics \cite{tsakalakos:08}. This deviation
from bulk behaviour needs to be taken into account at the time of describing the device operation mechanisms, a
requirement which may preclude the use of standard macroscopic device simulation models commonly
used in bulk photovoltaics. Similar issues are encountered in the field of NS-based light
emitting and amplifying devices, however, the regime of operation is inverted in SCs, since
the light needs to be trapped and the charge carriers are to be extracted. While light-trapping 
techniques have been successfully implemented, leading to a substantial efficiency enhancement, the
increase of collection efficiency remains a critical issue, mainly due to the strong
interaction of charge carriers with their environment and the large number of interfaces associated with the
NS. In this review, we will thus focus on SC devices where the NS
possess an electronic functionality as an electrically coupled absorber material, and where both
carrier generation as well as extraction are being considered. Due to the emphasis on high 
efficiencies rather than low cost, only concepts based on regular inorganic NS shall be considered. These two
restrictions exclude the vast and rapidly growing fields of computational tools for nanophotonic light management
\cite{atwater:10,mokkapati:12} on the one hand,  and for organic and hybrid SC \cite{wang:09,kanai:10} (which are both
also largely based on NS but cannot be counted among the high-efficiency concepts) on the other hand. Also, compared to
more established technologies, the concepts presented here have not in all cases been successfully implemented yet, and experimental device 
characteristics are thus of limited validity for a reliable verification of the models providing 
the theoretical predictions for the device performance.
\IEEEpubidadjcol

The review is organized as follows. In section II, the use of NS in the implementation of
different specific high efficiency concepts is discussed. In a third section, the requirements for
appropriate models specific to the related device structures are formulated. Based on these findings, different classes of models are
identified in section IV and positioned with respect to a general hierarchy of device simulation
approaches. Section V provides an overview of existing implementations for the most
important types of NS-based SC, namely those based on quantum wells (QW) and
quantum dots (QD) with both optical and electronic functionality. Section VI summarizes the main
remaining challenges and open problems in nanostructured PV device simulation and thus points at
potential directions of future developments. Section VII concludes the review.

\section{Nanostructure-based implementation of high efficiency solar cells}
The idea of utilizing low dimensional structures for novel high efficiency SC concepts dates
back to the early days of epitaxial semiconductor growth in the eighties \cite{wanlus:82,chaffin:84}, and
became increasingly popular after the group of Barnham at Imperial College put forward
the concept of the quantum well solar cell (QWSC) in the early nineties \cite{barnham:91}, which
still represents the prototype of a successfully implemented NS PV device. At the
beginning of the last decade, most of the concepts investigated today had been conceived
\cite{green:00}. In 2001, NS for photovoltaics was the topic of a landmark workshop at the MPI Dresden,
which, together with the accompanying publications, boosted both popularity and impact of this field
of research.  Two important representatives of these concepts with extensive utilization of
 NS - the multi quantum well/dot single junction SC (MQWSC/MQDSC) and the QW/QD
 superlattice (QWSL/QDSL) multi-junction SC  - shall briefly be reviewed below.

% Four important representative of these concepts with extensive utilization of
% nanostructures - the quantum well SC, the superlattice tandem, the intermediate band
% SC and the hot carrier SC - shall briefly be reviewed below.

%- review basic ideas of concept, including schematic drawing, show how nanostructures do the job,
%identify the most important physical mechanisms 

\begin{figure}[t]
\begin{center} 
\includegraphics[width=6cm]{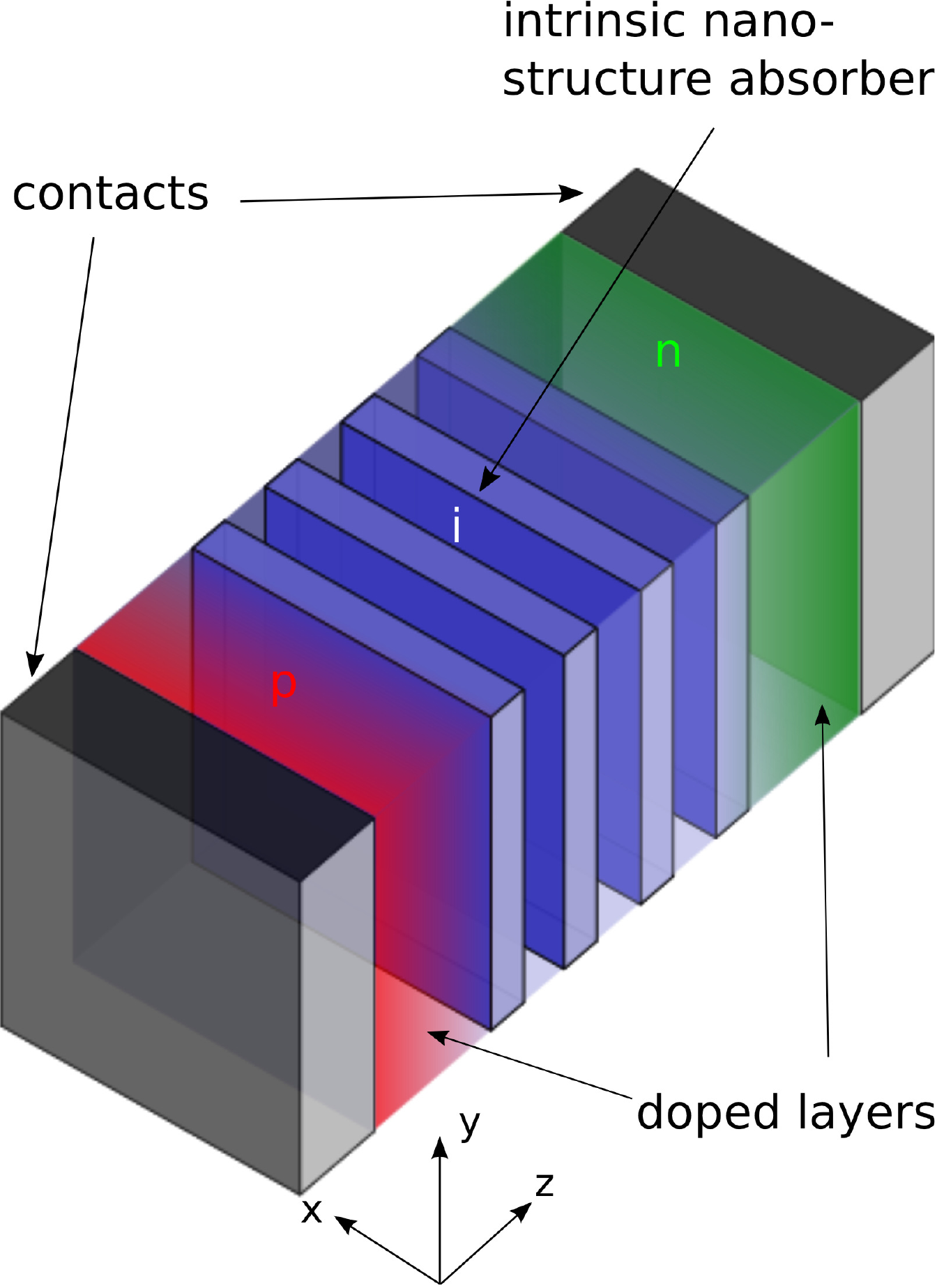}
\caption{Schematic representation of a bulk baseline p-i-n cell with
nanostructure absorbers inserted in the intrinsic region. While
photogeneration occurs in both bulk and NS components, carrier extraction
proceeds via the bulk states.\label{fig:mqwsc_struct}}
\end{center}
\end{figure}

\subsection{MQWSC/MQDSC - band gap engineering for single junction devices}
\begin{figure}[t]  
\begin{center}
\includegraphics[width=8cm]{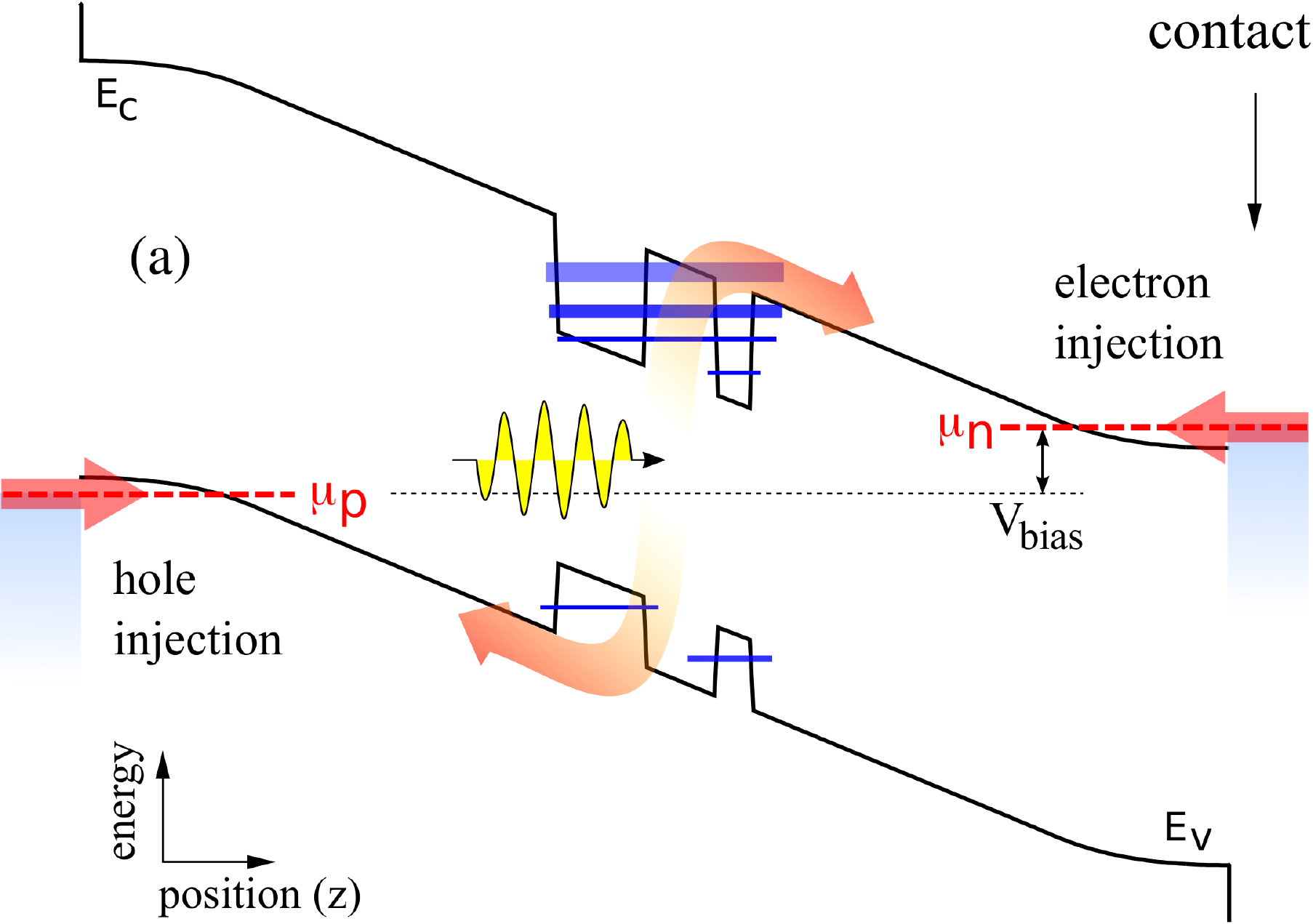} 
\end{center} 
\begin{center}
\includegraphics[width=2.8cm]{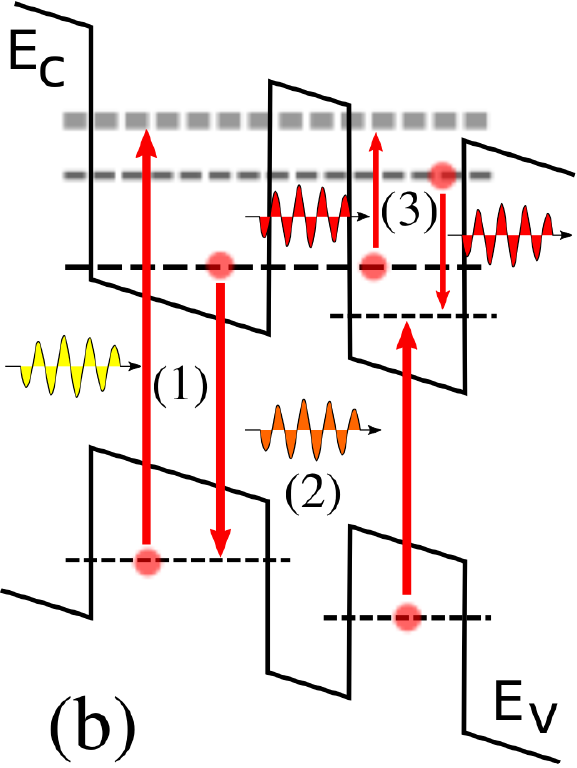}\includegraphics[width=2.8cm]{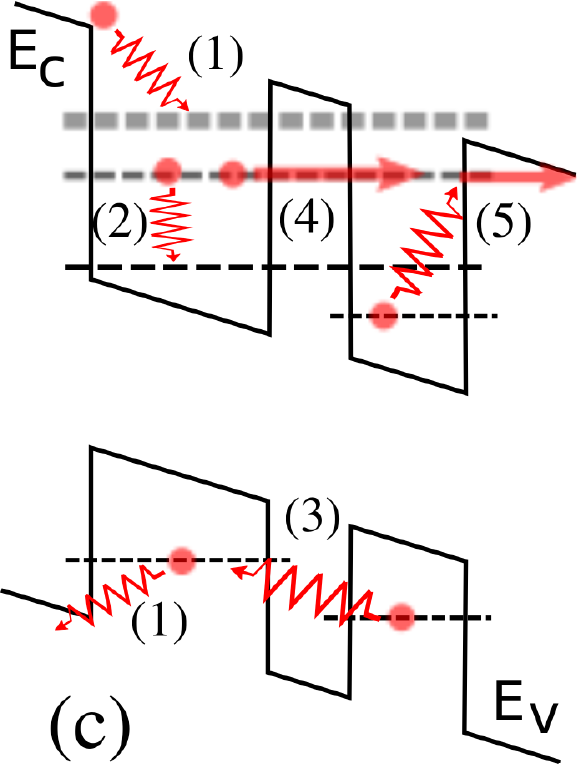}\includegraphics[width=2.8cm]{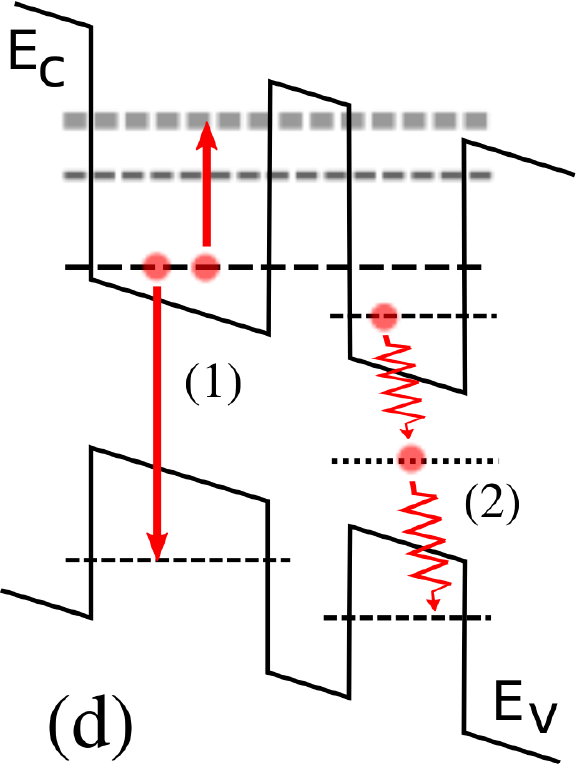}
\caption{(a) Schematic energy band diagram of a MQW/MQD solar cell.  The device (absorber/emitter)
states can be bound, quasi-bound or form a quasi-continuum.
At the contacts, photogenerated carriers are extracted and thermalized carriers are injected according to the chemical
potentials $\mu_{n,p}$ for electrons and holes, respectively, which are split by the voltage
$V_{bias}$. In the active device region, the physical processes relevant for the photovoltaic
operation are:
(b) radiative transitions, i.e. (1) interband photogeneration and radiative recombination as well as (2) radiative intraband transitions, which may also lead to (3) photon recycling; (c) coherent and dissipative quantum transport involving non-radiative intraband transitions, such as (1) phonon-mediated carrier capture and escape, (2) intraband relaxation, (3) scattering-assisted or (4) direct tunneling between	
absorber states, and (5) phonon-assisted carrier escape as combination of the two; (d) non-radiative recombination, via
(1) the Auger mechanism or (2) deep trap states.
\label{fig:mqwsc}}
\end{center}
\end{figure}

As one of the first NS-based high efficiency concepts being experimentally realized, the
QWSC now represents the most mature and industry relevant technology in
this field. The main idea of the concept is to extend the absorption range of a high band gap
material via the insertion of thin layers of lower band gap material in the intrinsic absorber
region of a single junction PV device, where the effective band gap is then determined by
the quantum well states. In the MQW concept (Fig. \ref{fig:mqwsc_struct}), the quantum wells are
separated by thick regions of high band gap material, and in order to contribute to the photocurrent, the charge
carriers generated in the quantum wells need to escape to the conduction band of the bulk material 
via thermionic emission or phonon-assisted tunneling (Fig. \ref{fig:mqwsc}). The
original hope for an independent optimizability of open circuit voltage $V_{oc}$ and short circuit
current $J_{sc}$ via the use of high and low band gap materials \cite{barnham:91} was soon 
shown to fail on thermodynamical grounds \cite{corkish:93,araujo:94,anderson:95}: the $V_{oc}$
was found to correspond rather to that of a bulk cell with an effective band gap determined by the
nanostructure states.  On the other hand, $J_{sc}$ could be effectively enhanced, as the escape
process proved to be highly efficient with probabilities close to unity at room temperature for effective barrier potentials of the order of 0.1-0.2  eV. The use of strain-balancing techniques allowed to overcome initial difficulties with strain-related dislocation formation \cite{ned:02_2}, such that dislocation free samples with up to 65 QW layers could be grown, and a maximum efficiency of 28.3\% could be achieved (under high concentration)\cite{adams:10}, which corresponds to the highest value for a NS-based single junction SC
architecture. There are similar approaches based on quantum dot layers \cite{hubbard:08,oshima:08}, 
which have the additional advantage of allowed radiative subband transitions, however, the QDs need 
either be large enough or strongly coupled in order to provide a quasi-continuum of subgap states. In contrast to the QW
devices, formation of dislocations and other surface defects, as well as achieving a high enough QD density for
sufficient absorption are still a critical issues in QD-based solar cell devices, and the resulting $V_{oc}$ loss is
higher.
   
 \begin{figure}[t]
\begin{center} 
\includegraphics[width=6cm]{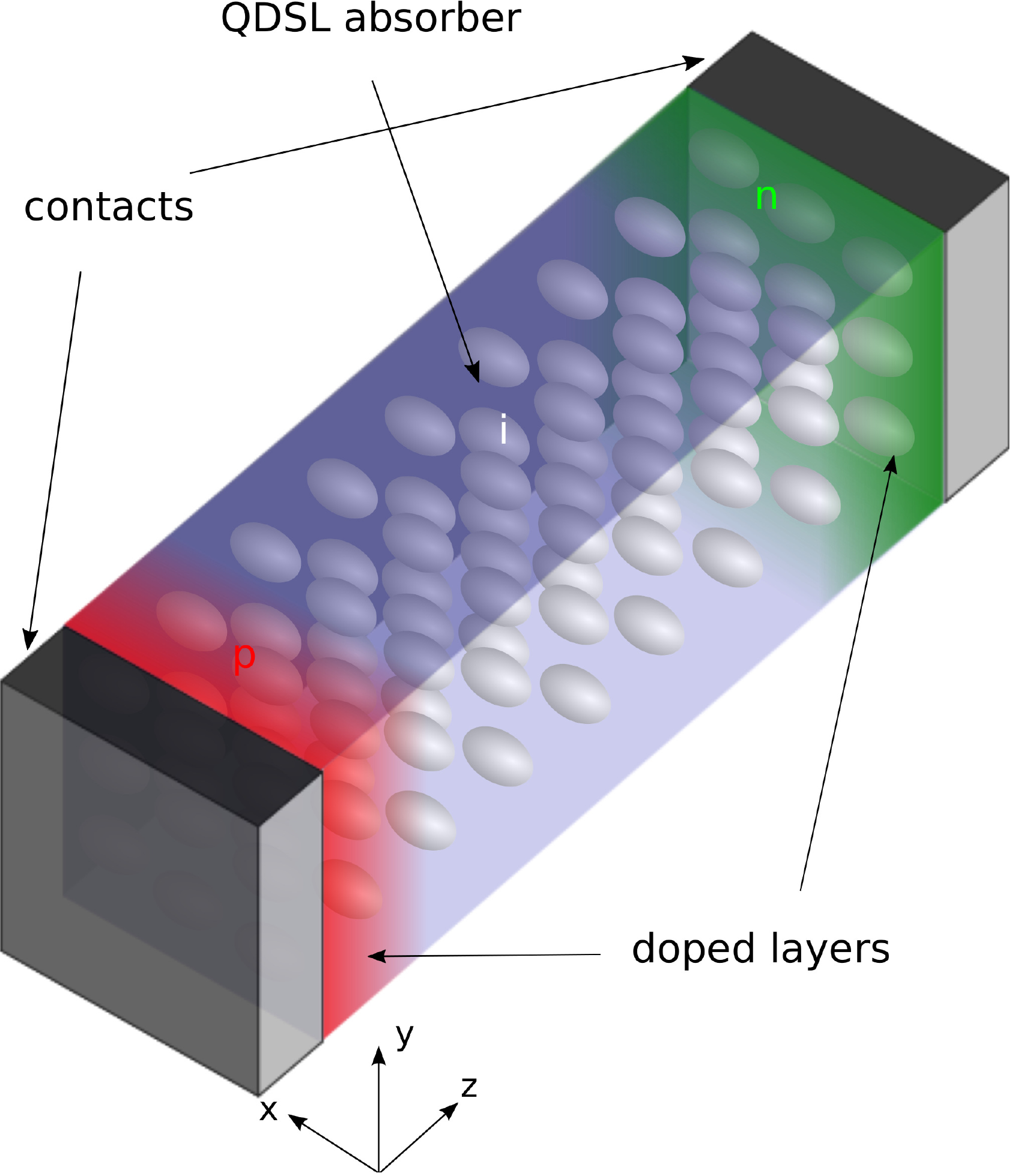}
\caption{Superlattice absorber component based on arrays of strongly
coupled QD. In contrast to the MQW device, both photogeneration and carrier
extraction proceed via nanostructure states\label{fig:qdsl_struct}.}
\end{center}
\end{figure}

\subsection{QWSL/QDSL-multi-junction SCs}
Multi-junction SC architectures rely on the vertical stacking of absorber materials with
decreasing band gap energies, which allows for a better utilization of the solar spectrum as
thermalization losses are reduced. Concentrator architectures with multi-junction cells
based on III-V semiconductor alloys yield record efficiencies up to 44\% \cite{green:13}. The
single component cells are connected in series via tunnel junctions, the voltages of the subcells are thus added, 
while the current needs to be matched over the whole absorber stack. This current matching condition 
puts severe limitations to the optimum combination of band gap values. Depending on the material system, it is difficult
or even impossible to find a bulk absorber with the required properties. In this situation,
superstructures of coupled quantum wells, wires or dots can be used to engineer absorber components with the desired
band gap (Fig. \ref{fig:qdsl_struct}). The coupling of the QW/QD gives rise to a delocalization of the confined states,
which for periodically arranged nanostructures results in the formation of so-called minibands with tunable band edges.
As opposed to the case of the MQW or MQD architectures, transport of charge carriers generated in miniband states is also 
mediated by the latter and not by the extended states of the host material. Due to the prominent position of crystalline silicon SC, 
a large amount of research was dedicated to the fabrication of all-silicon tandem SC with a superlattice of silicon nanocrystals in 
silicon-based dielectric matrix material as high-band gap absorber component
\cite{conibeer:06,conibeer:08}. However, while the tunability of the absorption edge is routinely
achieved, efficient charge carrier transport remains a true challenge, and there exists up to now no
working superlattice cell relying entirely on transport in miniband states.
To get a notion for the challenges associated with this concept, one first should remember that, as
a general fact, the coupling of NS exhibiting quantum confinement leads to a reduction of
the confinement in the dimension of the coupling. It is thus impossible to achieve at the same time
high band gap values and high carrier mobilities by using a quantum well superlattice where
confinement is only present in the transport direction. On the other hand, the delocalization of the
wave functions in the case of quantum dot superlattices is very sensitive to irregularities in shape
or position of the dots which are always present in real samples. The most advantageous architecture
would thus be composed of conducting filaments or nanowires with sufficient radial confinement,
avoiding the conflict between localization for band gap tuning and delocalization for charge carrier
extraction by separating of the respective dimensions. 
 
\begin{figure}[t]  
\begin{center}
\includegraphics[width=8cm]{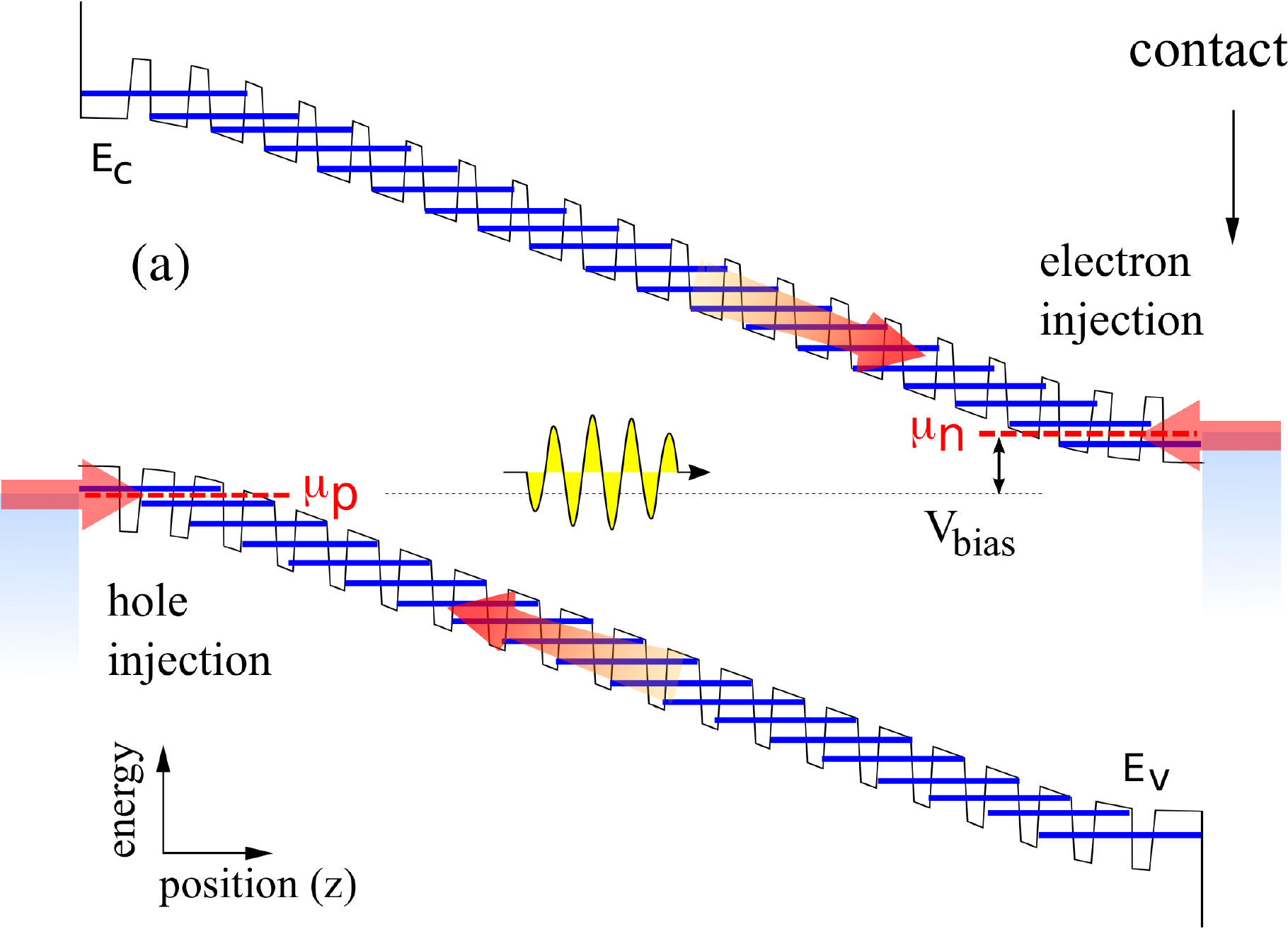} 
\end{center} 
\begin{center}
\includegraphics[width=2.8cm]{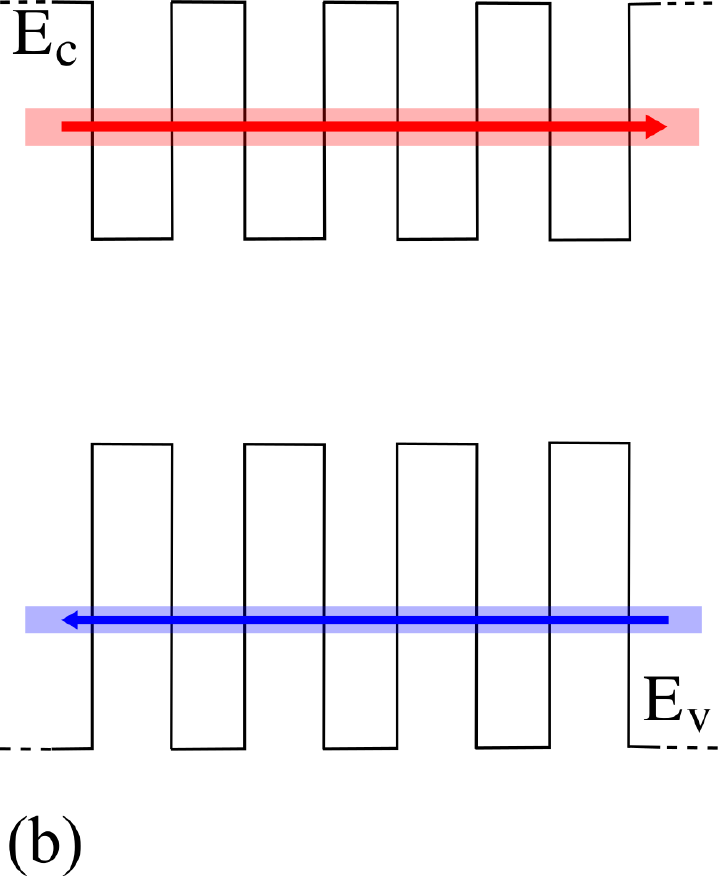}\includegraphics[width=2.8cm]{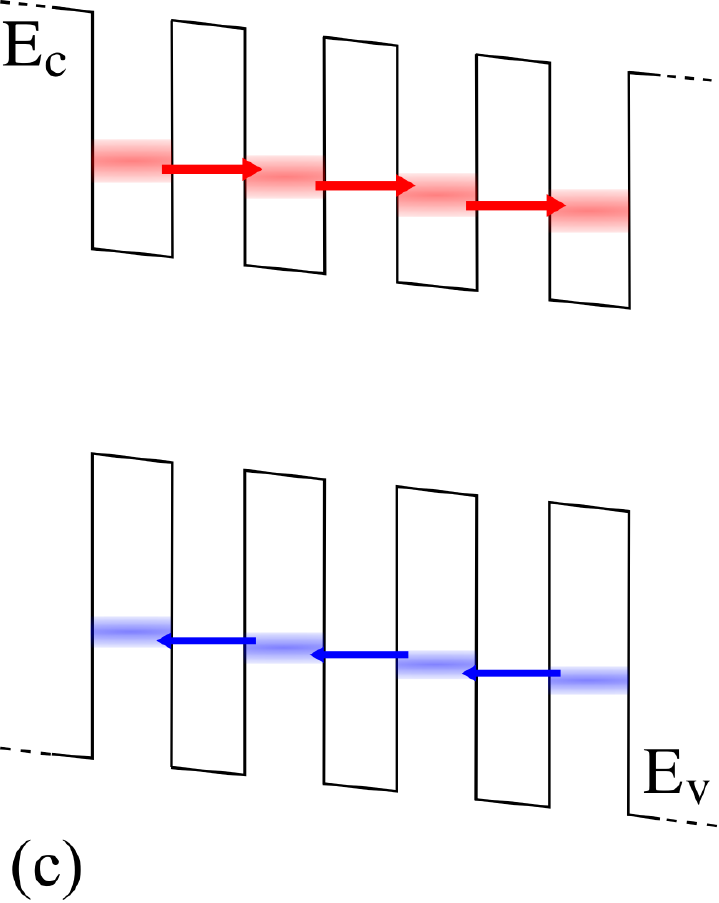}
\includegraphics[width=2.8cm]{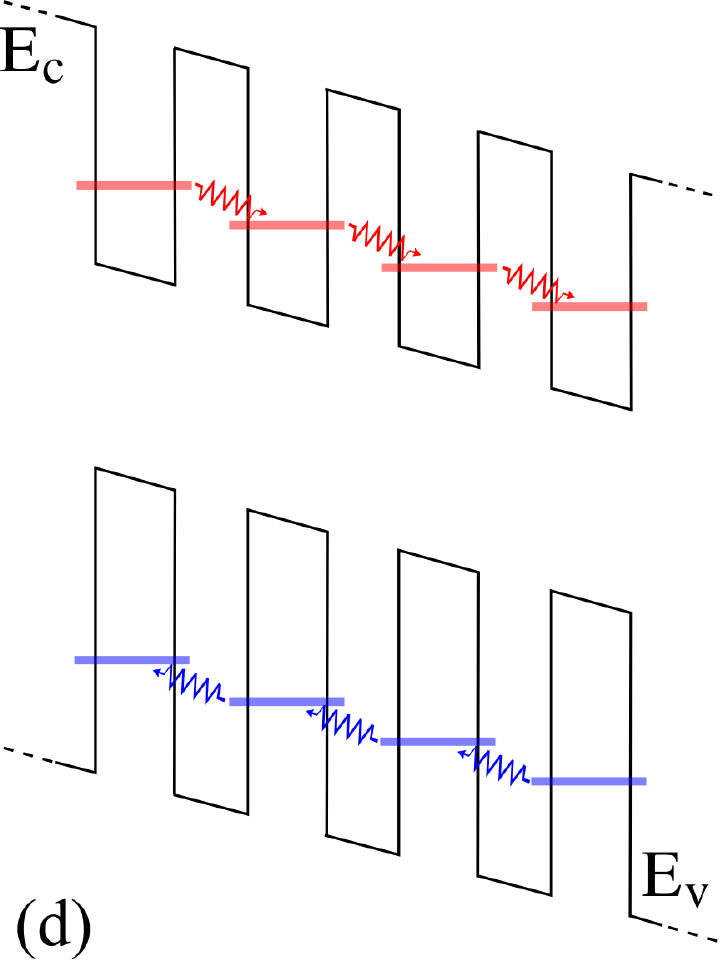}
\caption{(a) Schematic energy band diagram of a QWSL/QDSL solar cell. In this configuration, both
absorption and transport are mediated by the states of the strongly coupled nanostructures. 
Depending on the field in the absorber region, different transport regimes can be distinguished: (b)
miniband transport via fully delocalized states in the case of vanishing field; (c) sequential
tunneling via partially overlapping states in adjacent nanostructures at moderate fields; (d)
hopping transport in Wannier-Stark ladders assisted by inelastic scattering at strong fields.
\label{fig:qdsl}}
\end{center}  
\end{figure}

\section{Modelling requirements}
On a very fundamental level of description, a SC converts light into electrical energy. Any theory
suitable for the modelling of a SC device thus needs to be able to describe the fundamental
photovoltaic processes of charge carrier generation by photon absorption and charge carrier
extraction via 	carrier selective contacts, which usually involves transport. Even in ideal absorber
materials, the principle of detailed balance demands the presence of radiative recombination by
spontaneous and stimulated emission, and in the high-injection regime, the PV performance
of an ideal absorber is limited by the Auger recombination process, which is also
intrinsic, i.e., can not be avoided and needs thus be considered for any realistic estimate of limiting efficiency. 

Depending on the functionality of the NS within the SC device, some or all of
these processes are affected by the peculiar physical properties of the nanoscale component, with
major implications for the requirements on any suitable simulation approach. Let us consider in the
following the different stages of the photovoltaic energy conversion with focus on these modelling
requirements.
 
\subsection{Exciton generation by photon absorption}
The interaction of electromagnetic radiation with the electrons of a solid results in electronic
excitations, whose nature depends on the energy and polarization of the incident radiation field and
on the available electronic states. While this interaction is inherently nonlocal from the
electronic point of view, generation is usually determined based on locally defined absorption
coefficients that reflect all optical excitations possible at a given photon energy, weighted by
their respective strength. 

The main quantities determining the generation of excitons are, thus, the local value of the
transverse electromagnetic field on the one hand, and the local value of the dipole or momentum
matrix elements and of the density of occupied initial and empty final states of the optical
transitions, respectively, on the other hand. While the transverse EM fields are barely affected by semiconductor
NS in the range of few nm, the electronic quantities reflect the impact of
symmetry breaking and wave function localization due to the departure from bulk material towards significant spatial
variation of material properties on short length scales. There is, for instance, a
contribution of direct transitions in silicon quantum dots, which is absent in bulk and is
beneficial to the absorption \cite{allan:02,luo:11}, and also an increase in oscillator strength due
to larger overlap of electron and hole wave functions. Hence,
the optical modelling is not essentially different from that for bulk or conventional thin film
devices, but any suitable description of the light-matter coupling needs to reflect the nanoscale
extensions of the absorber.

\subsection{Charge transport to carrier selective contacts}
This is one of the most critical aspects in NS-based PV devices. Indeed, while the tayloring of
the absorption via band structure engineering was shown to work out for a range of SC devices based
on QW \cite{wagner:07} or QD \cite{zacharias:02}, the efficiency of carrier collection has
until now remained below that in the bulk counterparts, in some cases on a dramatically poor level, especially in devices where
transport proceeds via states with increased degree of localization. There are several reasons for the observed
performance issue, which are associated with different stages in the extraction process, such as
exciton dissociation, carrier escape and capture between localized and extended states, mobility issues associated with scattering and the
virtual absence of true miniband formation in realistic situation, which shall be discussed below. 

\subsubsection{Exciton dissociation}
Since electrons and holes have to be extracted via separate contacts, the dissociation of the
exciton and subsequent charge separation need to occur at some point in the device. In bipolar bulk
SC, exciton binding energies are usually in the range of a few meV, resulting in thermal
dissociation immediately after generation, and the generation can thus safely be described in terms
of non-interacting electron-hole pairs.
In isolated NS on the other hand, exciton binding energies can easily amount to multiples of $k_{B}T$
\cite{luo:11}, which can give rise to prominent excitonic features in the absorption and photocurrent spectra
\cite{paxman:93}, and can severely inhibit an efficient charge separation and even result in a
transport behaviour dominated by exciton diffusion, similar to the situation in organic SC
devices, where a specially designed bulk-heterojunction interface is required for exciton
dissociation. As a SC based on exciton diffusion is not likely to reach very high
efficiencies, the description of excitons in the simulation of high-efficiency SC devices is
primarily focused on the excitonic enhancement of optical transitions close to the effective band
edge.

\subsubsection{Carrier escape and capture} 
Basically, there are two types of NS-absorber-based SC devices. In the first
case, the NS act as mere absorbers, with the charge carriers generated in
localized states escaping to the extended states of the bulk host material via thermionic emission
or phonon-assisted tunneling (Fig. \ref{fig:mqwsc}c). A prototype example of this kind is the MQWSC
with large spacing between the quantum wells. In this situation, the charge carrier extraction is limited by the efficiency of the
escape process and by the probability of subsequent carrier capture in the lower-lying
localized states of NS encountered on the way to the contacts. A realistic assessment of
the device performance thus relies on an accurate determination of the associated rates.

\subsubsection{Carrier mobility and relaxation}
In the second type of NS-absorber-based solar cell device, the NS mediate not
only the absorption, but also the carrier transport to the contacts (Fig. \ref{fig:qdsl}a). These
states thus need to be to some degree extended, which amounts to the requirement of coupling between the single
NS. Indeed, most of the concepts of this kind that have been proposed are based on the
formation of minibands in superlattices of quantum wells or quantum dots (Fig. \ref{fig:qdsl}b).
However, in realistic situations, due to the presence of internal fields and any kind of spatial and compositional
disorder, the NS states usually show a high degree of localization, such that tunneling is
restricted to nearest neighbors (Fig. \ref{fig:qdsl}c). In addition to inducing localization, these
effects lead to a misalignement of energy levels in adjacent NS, with the formation of Wannier-Stark ladders in the extreme case of very
strong fields \cite{wacker:02}, in which situation transport is only possible via an inelastic
scattering process and is best described by hopping (Fig. \ref{fig:qdsl}d). Even in the ideal case
of a perfect miniband, the associated Bloch mobility may be critically low, as in the case of silicon QD in dielectric 
matrix material \cite{jiang:06}. In coupled QD systems, due to the sparsity of the DOS, the
presence of charge on the NS can lead to pronounced shifts in the energy level structure
\cite{luo:11}, inhibiting transport \mbox{($\rightarrow$ Coulomb} blockade regime). The bulk picture
of charge carrier transport in band states does thus in general not provide a proper description of the real
situation, which in addition to the remaining coherence and non-locality effects should include all the
localization effects and scattering mechanisms required to overcome energetic misalignement.
The modification of the electronic structure due to the presence of NS also affects the scattering processes
responsible for the limitation of mobility and for carrier relaxation, such as electron-phonon
interaction, which becomes especially relevant in devices where this form of energy dissipation
should be suppressed, as in the hot carrier concept, and in the case where extraction distances are
on the order of the mean free path and transport approaches the ballistic regime.

\subsection{Recombination}
The poor transport properties exhibited by many of the NS-based SC devices would
not be as detrimental as they are if recombination was limited to radiative processes, since radiative livetimes
are still long enough to allow for carrier extraction, even though they are reduced in NS as compared to the bulk 
(due to localization and local reduction of the effective band gap). However, there is an inherent increase in surface
area associated with the presence of NS, which are thus likely to act as centers of non-radiative recombination. 	In
some cases, it is possible to reduce the NS-related defects via specially engineered strain-balancing techniques \cite{ned:99_2}
or via passivation \cite{ding:icans11}, but in general, the insertion of NS leads to
losses in the open circuit voltage of the SC. Like for the generation process, a proper
consideration of recombination processes requires a careful consideration of the wave functions and local density of
NS states together with the occupation of these states in general non-equilibrium
conditions.

\section{Classification and hierarchy of models} 

Based on underlying assumptions, scope and predictive power, the majority of modelling approaches for
NS-based high-efficiency SC devices can be roughly categorized in three classes:\\
(i) Thermodynamic or detailed balance theories for ideal systems;\\
(ii) Models based on analytical or numerical solution of the macroscopic semiconductor transport equations. 
The source and sink terms for carrier generation and recombination are included together with equations for the 
rates of carrier exchange between localized and extended states (where separate microscopic models
for the localized states might be used);\\
(iii) Advanced quantum-kinetic models reflecting the microscopic mechanisms of the photovoltaic processes.

While the models in the first category rely on highly idealized assumptions and therefore primarily
provide upper limiting efficiencies, the latter two classes of approaches describe more realistic
situations, including an estimate of the relevant loss processes, and aim at reproducing real device characteristics.

\subsection{Thermodynamic and detailed-balance theories for ideal systems}

This type of limiting efficiency analysis dates back to the landmark
paper of Shockley and Queisser \cite{shockley:61}, where the current from a 
SC with only radiative recombination is calculated as the difference
between absorbed and emitted radiative flux, 
\begin{align}
j/q=\Delta \Phi_{\gamma},
\end{align}
making use of the principle of detailed balance for the computation of the latter
\cite{roosbroeck:54}. In its original form, this approach assumes infinite mobility corresponding to
a constant QFL separation, 	vanishing reflectivity, complete 	transparency for photon energies below
the band 	gap and complete absorption for photon energies above it.

Within the above setting of idealized conditions, an alternative starting point is a modification
of the ideal diode equation \cite{anderson:95}
\begin{align}
J(V)=&J_{0}[\exp(qV/kT)- 1]-J_{G}+J_{R}
\end{align}
where the generation and recombination currents $J_{G}$ and $J_{R}$ are composed of the
contributions of the bulk ''baseline'' cell and of the NS components. The
NS contributions are related to those of the baseline using correction factors for quantities such
as geometry (fraction of NS material), oscillator strength and
density of states (for radiative transitions).
 
\subsection{Macroscopic continuum and hybrid transport models}
The main shortcoming of ideal theories and global detailed balance approaches is the lack of
transport, i.e., the assumption of unit collection efficiency, which is often inappropriate in
NS-based SC devices. In order to introduce the effects of finite mobility, actual
transport equations need to be solved for the charge carriers. There, the main difficulty resides in
the consideration of the contribution of NS states with a higher degree of localization.
To obtain the self-consistent occupation of NS ($c$) and bulk host ($b$) states, separate
but coupled rate equations need to be solved for the densities of carriers occupying different types of states. Under
 the standard assumption of continuum and complete thermalization, this amounts to solving the
steady-state continuity equations, which for electrons read \cite{ramey:03}
\begin{align}
0=\partial_{t} n_{b}(\mathbf{r},t)=&\frac{1}{q}\nabla \cdot
\mathbf{j}_{n,b}(\mathbf{r})+G_{b}(\mathbf{r})-U_{b}(\mathbf{r})\nonumber\\
&+R_{n,c\rightarrow
b}(\mathbf{r})-R_{n,b\rightarrow c}(\mathbf{r}),\label{eq:dd_bulk}\\
0=\partial_{t} n_{c}(\mathbf{r},t)=&\frac{1}{q}\nabla \cdot
\mathbf{j}_{n,c}(\mathbf{r})+G_{c}(\mathbf{r})-U_{c}(\mathbf{r})\nonumber\\
&+R_{n,b\rightarrow
c}(\mathbf{r})-R_{n,c\rightarrow b}(\mathbf{r}),\label{eq:dd_ns}
\end{align}
with the currents of drift-diffusion type
\begin{align}
\mathbf{j}_{n,i}(\mathbf{r})=&q\mu_{n,i}(\mathbf{r})n_{i}(\mathbf{r})\boldsymbol{\mathcal
E}(\mathbf{r})+qD_{n,i}(\mathbf{r})\nabla n_{i}(\mathbf{r}), ~(i=b,c),
\end{align}
where $\mu$ denotes the mobility, $D$ the diffusion coefficient, $\boldsymbol{\mathcal{E}}$ the
electric field, and $q$ the elementary charge. Similar equations have to be solved simultaneously for the hole
densities $p_{b,c}$ and currents $\mathbf{j}_{p,i}(\mathbf{r})$ ($i=b,c$), respectively. In the
above equations, the generation rate $G_{i}=G_{i}[\Phi_{\gamma}(\mathbf{r}),\alpha_{i}(\mathbf{r},E_{\gamma})]$ ($i=b,c$)
is a functional of the local photon flux and the local absorption coefficient, and the recombination 
rate $U_{i}=U_{i}[n_{i},p_{i},\ldots]$ ($i=b,c$) depends on the carrier densities and on
recombination-mechanism (radiative, Shockley-Read-Hall, Auger) specific parameters. In the case
where, due to strong localization, direct transport between NS states can be neglected, the current term in
Eq. \eqref{eq:dd_ns} is absent, and exchange of carriers between localized states always proceeds
via extended states. The coupling of the equations for localized and
extended states is provided via the escape and capture rates on the second line of Eqs.
\eqref{eq:dd_bulk} and \eqref{eq:dd_ns}, which are subject to the detailed balance condition and
are commonly expressed via the density in the initial state and an associated lifetime, e.g.,
\begin{align}
R_{n,c\rightarrow b}=\frac{n_{c}}{\tau_{n,c\rightarrow b}}.
\end{align}
The lifetimes (which are local quantities) are either used as fitting parameters to reproduce
experimental characteristics, or derived from an appropriate description for the microscopic mechanisms of 
the scattering process responsible for the coupling, e.g., via Fermi's Golden Rule, based on the
solution of the Schr\"odinger equation for the NS states and energies. For consistency, the absorption
coefficient and mobilities used in the generation term and the current expressions should be
computed on the basis of the same solutions of the microscopic equations for the electronic structure.

Finally, the transport (and Schr\"odinger) equations have to be solved (self-consistently) together
with Poisson's equation for the spatial variation of the electrostatic potential $\phi$ ($\boldsymbol{\mathcal{E}}=-\nabla \phi$),
\begin{align}
\nabla\cdot[\varepsilon(\mathbf{r})\nabla \phi(\mathbf{r})]=q n_{tot}(\mathbf{r}), \label{eq:poisson}
\end{align}
with $\varepsilon$ denoting the static dielectric function and the total ($tot$) charge density  given by the sum of contributions from localized and extended
states,
\begin{align}
n_{tot}=&N_{d}+n_{b}-p_{b}+n_{c}-p_{c},
\end{align} 
where $N_{d}$ is the net charge density due to ionized dopants.

The resulting \emph{hybrid} (macroscopic - microscopic) approach, which represents a kind of ''poor
man's'' multi-scale modelling, can be used to reproduce experimental device characteristics with
remarkable accuracy. However, being a macroscopic and local model, any situation requiring energy
resolution, non-locality or coherence cannot be described properly, e.g., non-thermalized carrier
distributions or resonant tunneling. To include such processes in a consistent description, a truly
microscopic picture of carrier transport is to be used.

\subsection{Microscopic quantum-kinetic theories}
\begin{figure}[t] 
\begin{center} 
\includegraphics[width=8cm]{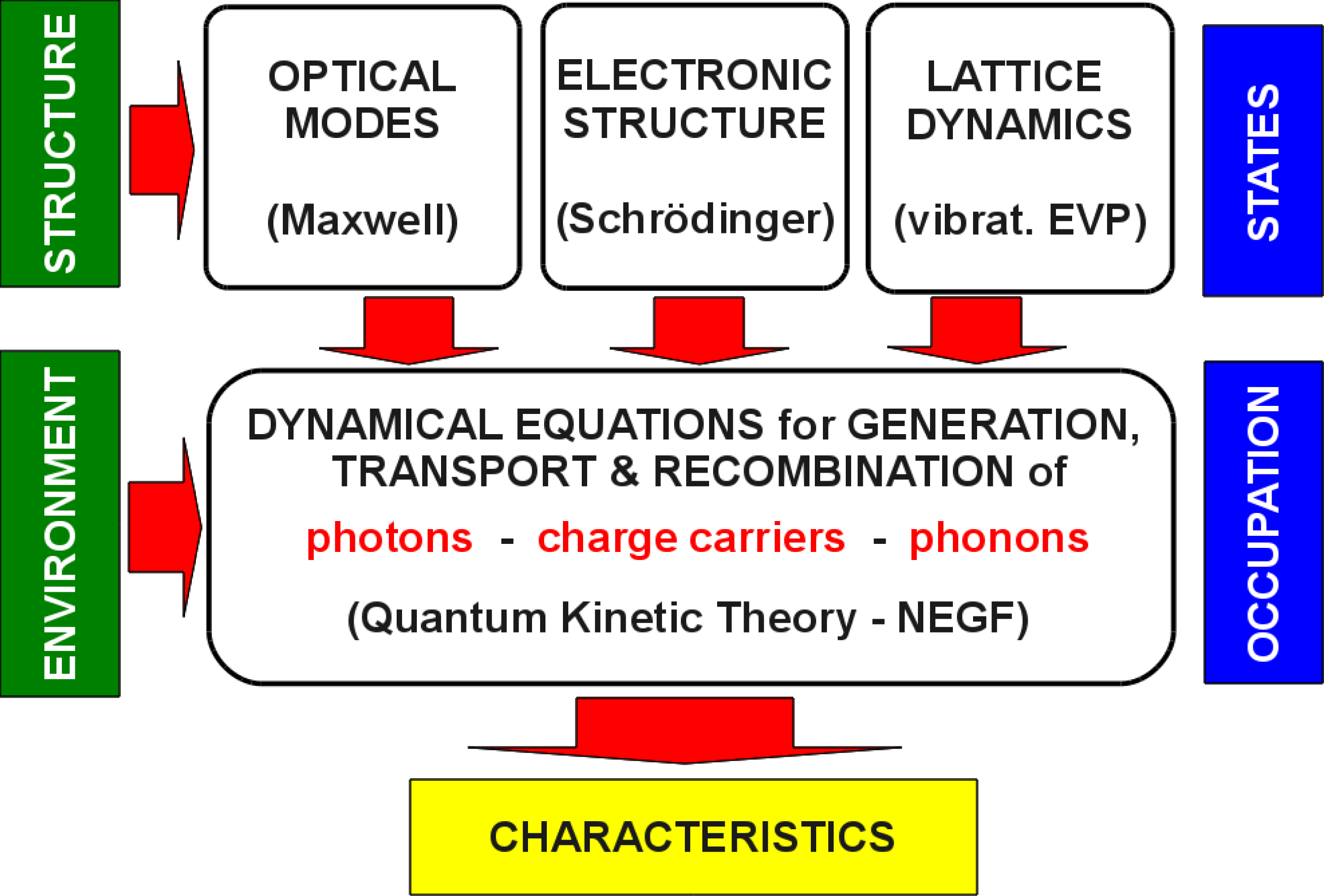}
\caption{Multi-scale \& multi-physics framework for a mesoscopic quantum-kinetic theory of
nanostructure-based photovoltaic devices. Based on the macro- and microstructure of the functional
components of the device, the relevant optical, electronic and vibrational states of the open system
are determined by the solution of corresponding eigenvalue problems (EVP). The occupation of these
states in response of the given environmental conditions (illumination, applied voltage,
temperature, etc.) is determined via the solution of dynamical equations encoding the physical
processes of generation, transport and recombination. 
\label{fig:concept}}
\end{center}
\end{figure}

The challenge in using a microscopic approach resides in the fact that many modelling requirements
that were automatically met by the macroscopic approach, such as, e.g., open boundary
conditions, carrier relaxation and non-equilibrium occupation, are very hard to satisfy in a
quantum-mechanical picture and have to be addressed explicitly on the basis of scattering states,
which results in an adequate description to be found only at the quantum-kinetic level.
There, the non-equilibrium Green's function formalism (NEGF), popular in nano-electronics \cite{datta:95,datta:05} and quantum
optics \cite{haug:96,schaefer:02,haug:04}, provides an ideal framework for the formulation of a
comprehensive quantum theory of NS-based PV devices \cite{ae:jcel_review}.
The steady-state photovoltaic balance equation in \eqref{eq:dd_bulk} is replaced by its
microscopic quantum-kinetic counterpart \cite{kadanoff:62,keldysh:65}
\begin{align}
\nabla\cdot 
\mathbf{j}&(\mathbf{r})
=-\frac{2}{V}\int\frac{dE}{2\pi\hbar}\int d^3 r'\Big[\Sigma^{R}
(\mathbf{r},\mathbf{r}';E)G^{<}(\mathbf{r}',\mathbf{r};E)
\nonumber\\&+\Sigma^{<}(\mathbf{r},\mathbf{r}';E)G^{A}(\mathbf{r}',\mathbf{r};E)
-G^{R}(\mathbf{r},\mathbf{r}';E)\Sigma^{<}(\mathbf{r}',\mathbf{r};E)\nonumber\\&-
G^{<}(\mathbf{r},\mathbf{r}';E)\Sigma^{A}(\mathbf{r}',\mathbf{r};E)\Big]\label{eq:currcons}
\end{align}
in terms of Green's functions $G$ and self-energies $\Sigma$ for charge carriers, the right hand
side of the above equation thus represents a general expression of the total scattering rate
(intra- and interband). The Green's functions as the basic quantities are determined by the
steady-state Dyson and Keldysh equations (omitting the fixed energy argument $E$)
\begin{align}
\int &d \mathbf{r}_{1}\Big[\big\{G_{0}^{R}\big\}^{-1}(\mathbf{r},\mathbf{r}_{1})
-\Sigma^{R}(\mathbf{r},\mathbf{r}_{1})\Big]G^{R}(\mathbf{r}_{1},\mathbf{r}')=~\delta(\mathbf{r}-\mathbf{r}'),\\
&G^{\lessgtr}(\mathbf{r},\mathbf{r}')=\int d\mathbf{r}_{1}\int d
\mathbf{r}_{2}G^{R}(\mathbf{r},\mathbf{r}_{1})\Sigma^{\lessgtr}(\mathbf{r}_{1},\mathbf{r}_{2})
G^{A}(\mathbf{r}_{2},\mathbf{r}').
\end{align} 
While the retarded and advanced Green's functions $G^{R/A}$ are related to the density of states
(DOS), the correlation functions $G^{\lessgtr}$ additionally contain information on the
(non-equilibrium) occupation of these states. The non-interacting system is described by $G_{0}$. The self-energies
$\Sigma$, on the other hand, are scattering functions describing the renormalization of the Green's
functions due to coupling to the the environment, in the form of interactions with photons 
($\rightarrow$ photogeneration, radiative recombination), phonons ($\rightarrow$ relaxation,
indirect transitions) and other carriers ($\rightarrow$ excitons, Auger processes). An additional
self-energy term describes injection and extraction of carriers a contacts with arbitrary chemical
potential, 	enabling the treatment of an open non-equilibrium system. The macroscopic photovoltaic
device characteristics are obtained from the Green's functions via the respective expressions for
steady-state carrier and current density, which for electrons read
\begin{align}
n({\mathbf r})=&-i\int \frac{dE}{2\pi}G^{<}({\mathbf r},{\mathbf r};E)
 \label{eq:steaddens}
  \end{align}
and  
  \begin{equation}  
\mathbf{j}_{n}(\mathbf{r})=
\lim_{\mathbf{r'}\rightarrow\mathbf{r}}\frac{\hbar}{2m_{0}}[\nabla_{\mathbf{r}}
-\nabla_{\mathbf{r'}}]\int \frac{dE}{2\pi}G^{<}(\mathbf{r},\mathbf{r'};E),
\end{equation}
respectively. Similar expressions exist for the holes in terms of $G^{>}$.
As for the models based on the macroscopic transport equations, the computation of the Green's
functions needs to be coupled self-consistently to the determination of the electrostatic potential
from Poisson's equation, using the above expressions for the carrier densities in Eq.
\eqref{eq:poisson}. Similar sets of equations can also be formulated for the optical and vibrational
degrees of freedom (i.e., for photons and phonons), which then provides the desired comprehensive microscopic
theory of NS-based optoelectronic devices (Fig. \ref{fig:concept}). For a more in-depth discussion
of the NEGF formalism as applied to novel SC architectures, which is out of the scope of this review, 
the reader referred to \cite{ae:jcel_review} and references therein.

\section{Overview of existing modelling approaches}
Here, we review the most relevant literature on theoretical descriptions and modelling approaches
for specific types of NS-based high efficiency SC, with focus on models that
consider not only a single step of the photovoltaic process, but which can actually be used to
obtain the current-voltage characteristics - and with that the efficiency - of the device. This
strongly reduces the amount of available literature in contrast to the huge number of papers on the
optoelectronic properties of QW and QD systems.

\subsection{Multi-QW/QD solar cell}
This type of SC comprises the approaches where NS extend the absorption range of
the bulk device, but generated charge needs to escape to the bulk states in order to contribute to
the photocurrent, such as, e.g., in the case of MQWSC or MQDSC\footnote{See \cite{ae:thesis} and \cite{ae:bookchapter_nova} 
for a more extensive discussion of the physical foundations, theoretical description and numerical
simulation of QWSC.}. 

\subsubsection{Thermodynamic models and ideal theories}
The earliest models for the MQWSC were thermodynamical theories of ideal devices. The original Shockley-Queisser
formalism for single junction bulk devices was extended by Henry  \cite{henry:80} to the case
of multiple band gaps. The detailed balance approach of Corkish and
Green \cite{corkish:93} treated the QW as an incremental cell in addition to the
baseline high bandgap bulk cell, but without any coupling between the two and in a field-free
limit. $V_{oc}$ and $I_{sc}$ where then obtained from the
superposition of baseline and incremental cell. Ara\'ujo and Mart\'i
\cite{araujo:94} generalized the detailed balance analysis further, taking into account the
light path in the device, variation of refractive indices and the angular range
of incident and emitted radiation, and showing that for constant quasi-Fermi level separation
(QFLS), the emissivity equals the absorptivity, with the consequence that within this limit, QWSC could not
exceed the efficiency of an ideal homojunction device. The model of
Bremner, Honsberg and Corkish \cite{bremner:99}, based on ideas proposed
by Kettemann and Guillemoles \cite{kettemann:95}, allowed for QFLS
variations under assumption of radiative transitions between the different
levels, which yields considerable efficiency increase, but might not be applicable to the case of 
QWSC, due to the very small intraband transition matrix elements for in-plane polarization.
The origin of the variation in QFLS is not contained in the model, i.e. the
QFLS-step is not an emergent feature of the theory. A simple and intuitive ideal QWSC model was presented by Anderson
\cite{anderson:95}: his approach, which in philosophy is similar to the model
of Corkish and Green \cite{corkish:93}, is based on the ideal diode
current-voltage characteristics for bulk homojunctions, with the quantum well
material accounted for by enhancement factors for oscillator strength and DOS,
providing the modifications of generation and recombination. However,  the effects of QWs
on the transport properties are not considered. This model was further developed by different authors \cite{rimada:01,
lade:04,rimada:05,rimada:07}, including interface recombination and a better description of the QW absorption, to the
point where it could be used to fit experimental data.
 
 \subsubsection{Macroscopic continuum and hybrid transport models}
The first quantitative theory for a MQWSC beyond thermodynamical and
detailed balance approaches was developed by the group at Imperial College in the years after the
first experimental implementation of the concept, and consists of a model for the escape of
photogenerated carriers from QW \cite{nelson:93}, a semi-analytical model for
the entire spectral response \cite{paxman:93}, and a numerical model for the dark current \cite{nelson:94}, with
further improvements made in \cite{nelson:99}. In the approach, whose early stages are summarized in \cite{nelson:95},
dark currents are obtained by analytical or numerical solution of the electron and hole drift-diffusion equations including the
terms 	for generation and recombination and the coupling to Poisson's equation. The carrier density
is expressed in terms of the corresponding quasi-Fermi levels that are obtained from the transport
equations, and which are assumed to be conserved across the interface between barrier and well
material. In the case of QW, the expression for the density is modified by an additional factor to
adjust to the two-dimensional DOS calculated from the solution of the effective mass equations in
the envelope-function approximation \cite{bastard:86} providing the subband energies. To include
nonparabolicity of the light-hole band, a 4-band Kane model of the valence band is converted into
corresponding 1D effective mass equations for each carrier-type, with the effective mass acquiring
an energy dependence. The equations are solved numerically using a transfer-matrix method, which in
addition provides the transmission function of the confining barrier. The DOS calculated in this way
is also used to obtain the QW absorption that provide the generation rate. Excitonic contributions
to absorption are included via parametric models with dimensionality parameters for exciton binding
energies and oscillator strengths; the parameters are obtained from a fit to the solution of the
effective mass equation for excitons. The absolute excitonic absorption is scaled and convolved with
a Lorentzian for homogeneous broadening to fit the experimental data. Layer widths, composition and
doping 	levels are determined from growth record and characterization studies, and the minority carrier diffusion lengths
are calculated from layer doping and alloy fraction. The recombination rates are
determined by the bulk and QW densities and the recombination times including
radiative, SRH and Auger recombination, are obtained from fits to the
corresponding bulk control cell dark currents. The escape lifetime model describing carrier escape
from QW includes thermionic emission and (thermally assisted)
tunneling. The lifetime is derived from the escape current, which in
turn is determined by the carrier density at a fixed energy, given by the DOS
and the occupation, and the transmission function of the confining barrier at
that energy. At room temperature and moderate fields, the probability of
escape from QWs in the intrinsic region is set to unity, in accordance with
carrier escape experiments \cite{nelson:93}. The models were eventually combined by Conolly
\cite{conolly:00} to a comprehensive modelling approach for QWSC, using a simplified approach for the dark current and
assuming unit escape efficiency.

In spite of its comprehensiveness, there are several shortcomings in the approach, which in the
reproduction of experimental characteristics are in part compensated by adjusting a number of
fitting parameters. Most notably, the description of quasi-bound states close to the top of the well is
poor, since near the top of the wells, the envelope-function approximation breaks down, and the
QW-DOS gradually becomes bulk-like, which is not reflected in the model. While absorption and
emission are primarily dominated by the states close to the effective band edge. i.e., deep in the
well, these high-lying states play a decisive role in the escape and capture of carriers.
Furthermore, the DOS and correspondingly the absorption above the wells is assumed to be that of
homogeneous bulk, which is not the case due to the existence of quasi-bound states and higher
resonances. The assumption of unit escape breaks down for deeper wells, and capture processes are
not considered 	explicitely in the model. The model in its standard formulation does also not
include the effects of 	finite fields across the QWs, which lead to hybridization of QW and bulk
states, a shift of confinement levels (quantum confied Stark effect) and modified absorption due to
decreased 	overlap of asymmetric electron and hole wave functions. A further drawback is the
requirement for the assumption of  constant QFL separation in wells and barriers, which might not
hold in all cases, as indicated by experimental observations \cite{nelson:97} and numerical
investigations 	with spatially varying QFLS \cite{corkish:97,honsberg:99}.

Apart from the work by the Imperial College group, further modelling efforts include the model of
Varonides \cite{varonides:02}, which explicitly takes into account thermionic emission and tunneling
in a way very similar to Nelson \cite{nelson:93}, but tunneling is restricted to the triangular barrier of an isolated well. A more
comprehensive self-consistent Schr\"odinger-Poisson-drift-diffusion model for carrier generation,
recombination and transport in QWSC was developed by Ramey and Khoie \cite{ramey:03}. In distinction
to the above modelling framework, this approach describes also carrier capture into QW, solving
separate equations for densities of carriers in localized and extended states, and the escape model
considers the 2D-DOS, the subband energy level structure including valence-band mixing and escape 
from direct and indirect valleys, and the field and temperature dependence, but neglects tunneling 
escape, since it is assumed to be suppressed at room temperatures and low fields. The QW-absorption
is obtained from a semiempirical model \cite{lengyel:90}, not considering excitons, bandstructure or
field effects, and only non-radiative recombination (SRH) is described.
A more recent approach by Kailuweit et al. \cite{kailuweit:10} uses a commercial device simulator
for the transport, where barrier and well regions are treated as bulk materials (implicitly assuming
unit escape probability), together with the QW absorption model introduced in \cite{paxman:93}. For MQDSC, a simple semi-analytical
model in the spirit of Paxman \cite{paxman:93} is presented by Aroutiounian et al. in \cite{aroutiounian:01}
and extended later by the same authors to include emission and capture rates \cite{aroutiounian:05}.

While the models reviewed so far did not consider the (resonant) coupling of
multiple quantum wells (while still extracting carriers via the states of the bulk host), which can
significantly enhance carrier escape \cite{zachariou:98}, the inclusion of (coherent) multi-barrier
tunneling into 	the analysis of QWSC performance was accounted for in the model by Mohaidat et al.
\cite{mohaidat:94}, in which a numerical solution of the time dependent Schr\"odinger equation is
used to calculate resonant tunneling transport of photogenerated carriers in MQW with thin barriers.
However, the model was never embedded into a more comprehensive picture including explicitly carrier
generation, recombination and escape channels other than via tunneling.

\subsubsection{Mesoscopic quantum-kinetic approaches}
As mentioned above, the assumption of unit thermal escape
efficiency starts to break down for deep or strongly coupled QW. The necessary inclusion of
alternative escape channels requires consideration of the actual non-equilibrium occupation of \emph{quasi-bound} states as established under illumination and bias,  which is beyond the capabilities of the hybrid
approach. The formulation of a comprehensive quantum-kinetic theory for QWSC by Aeberhard and Morf 
\cite{ae:prb_08,ae:eupvsec09,ae:solmat_10,ae:spie10}, based on the non-equilibrium Green's function formalism, 	allowed
for the first time for a consistent description of carrier photogeneration, escape, capture and radiative recombination
mediated by states 	of arbitrary degree of localization resulting from realistic potential profiles and under arbitrary
non-equilibrium conditions. 	As the approach does not rely on local quasi-Fermi levels, 	it does not make any 
assumptions on 	the equilibration between nanostructure and bulk host states. In the description of escape and capture 	
processes, transport at any energy is considered, covering both coherent and phonon-assisted tunneling contributions. 	
However, due to the considerable numerical cost of the approach, only simple effective 	mass or 	few-band tight-binding 
models were used for the electronic structure, and characteristics could only be 	obtained for structures	 of mesoscopic spatial extension.

\subsection{QW/QD-superlattice solar cell} 
This category of solar cell comprises any concept based on miniband transport, e.g., by
QW or QD superlattices, often proposed as tunable absorber components in multi-junction
architectures. A basic ingredient of most models is the miniband structure associated with
delocalized superlattice states. For QWSL, approaches ranging from simple transfer
matrix formalisms \cite{courel:12} to advanced ab-initio methods \cite{kirchartz:09_SL} have been
used for this purpose. For QDSL, a popular procedure consists in the superposition of solutions to
the 1D Kronig-Penney model for a single band effective mass Hamiltonian \cite{jiang:06},
based on \cite{lazarenkova:01}. While the absorption (and, via detailed balance, the radiative
dark current) is determined in many cases from the computed electronic structure, the latter
does either not enter the transport problem at all \cite{courel:12,courel:12_JAP}, or only via an
effective density of states and a so-called \emph{Bloch mobility} 
\cite{kirchartz:09_SL} derived from the curvature in transport direction of the miniband for the
perfect superlattice in the absence of disorder and fields, together with an effective scattering lifetime \cite{jiang:06}.
Photogeneration and photocarrier transport for more realistic situations concerning internal fields
and inelastic scattering processes (electron-phonon interaction) were investigated using the
quantum-kinetic approach in \cite{ae:nrl_11} for Si-SiO$_{x}$ QWSL absorbers, confirming the
detrimental effect of high barriers on mobility and of thin barriers on confinement, and in
\cite{ae:oqel_12} for Si-SiC-SiO$_{x}$ QDSL solar cells, revealing strong QD wave function
localization already in moderate built-in fields.

\section{Remaining challenges and open problems}

The NS-based implementations of novel high efficiency solar cell concepts are of similar
complexity as state-of-the-art devices in nanoelectronics and solid-state lighting. However, advanced theoretical 
concepts for electron transport at the nanoscale are only slowly being adopted to simulate
nanostructure-based solar cells, although such concepts are, for example, readily applied in the simulation of
solid-state lighting”. This is in part due to the problem of scales: even under application of advanced light-trapping
schemes, 	the thickness required to absorb a substantial fraction of the incident light is still of the order of a few
hundred nm, which is beyond the capabilities of any model with atomistic resolution, 	especially in the case of
irregular, non-periodic arrangements of NS exhibiting 3D confinement, such as realistic QD arrays. 	The development of
adequate forms of \emph{multi-scale} modelling\cite{aufdermaur:11_TED}, ranging from parametrization of effective 
Hamiltonians from first-principles to the derivation of current- and charge-conserving interface conditions between
micro- and macroscopic transport models, will thus be a major focus in future research on simulation models for PV devices based on
NS. Due to the impact of nanoscale size on any kind of physical properties of the
NS-based device on the one hand, and the nature of the solar cell as an
optoelectronic device operating at elevated temperatures on the other hand, the multi-scale aspect
is accompanied by the notion of a \emph{multi-physics} approach that is required for an adequate description of the
device operation, i.e., including transport of charge, light and heat at all scales. 

Concerning the mesoscopic quantum-kinetic ingredient to the modelling framework, defect-mediated
processes and advanced electron-electron scattering mechanisms required to describe 
impact-ionization, Auger recombination or hot-electron distributions are still awaiting
implementation. In some cases of strong coupling, the single-particle picture may no longer be
appropriate, and quasi-particles such as excitons, polarons, plasmons or polaritons should be
considered. Finally, in realistic devices, an additional challenge arises from the presence of
disordered materials such as amorphous phases, e.g., in embedding matrix components, which requires
tedious configuration averages in order to extract representative device characteristics.

\section{Conclusions}

In this paper, theoretical approaches for the simulation of NS-based high-efficiency
solar cell devices have been reviewed, with focus on solar cells based on QW and QD absorbers with
varying degree of coupling between the NS components. In many cases, the effects of
quantum confinement on optical, electronic and vibrational properties of the NS interdicts the 
use of conventional bulk approaches. While the optical properties can be considered by explicitly
computing the absorption coefficient from the modified electronic structure and applying the
principle of detailed balance for the emission, the inclusion of the effects on transport and
non-radiative recombination requires a careful consideration of the microscopic mechanisms that couple 	
NS and bulk host states. If both tunneling and scattering processes are essential for the
device operation, as in the case of strongly coupled NS subject to significant electric
fields, an adequate description of photocarrier generation and extraction requires the use of
advanced microscopic quantum-kinetic approaches. Due to the huge computational cost of the latter, 
future approaches should aim at a combination of mesoscopic quantum-kinetic descriptions for the 
NS components and a macroscopic picture for the bulk regions in a general multi-scale and
multi-physics modelling framework.

% or


% \appendices
% % you can choose not to have a title for an appendix
% % if you want by leaving the argument blank
% \section{}
% % use section* for acknowledgement
% \section*{Acknowledgment}

\ifCLASSOPTIONcaptionsoff
  \newpage
\fi

   
 
\bibliographystyle{IEEEtran}
%\bibliography{/home/aeberurs/Biblio/bib_files/bandstructure_kp,/home/aeberurs/Biblio/bib_files/negf,/home/aeberurs/Biblio/bib_files/recombination,/home/aeberurs/Biblio/bib_files/pv,/home/aeberurs/Biblio/bib_files/qwsc,/home/aeberurs/Biblio/bib_files/qd,/home/aeberurs/Biblio/bib_files/hot_carriers,/home/aeberurs/Biblio/bib_files/sinova,/home/aeberurs/Biblio/bib_files/aeberurs,/home/aeberurs/Biblio/bib_files/scqmoptics,/home/aeberurs/Biblio/bib_files/multiscale_modelling}

 %\begin{thebibliography}{}
%    
\input{jstqe_review_final_clean.bbl}

 %\end{thebibliography}
 
%\begin{biography}[{\includegraphics[width=1in,height=1.25in,clip,keepaspectratio]{mshell}}]{Michael Shell}

\begin{IEEEbiography}[{\includegraphics[width=1in,height=1.25in,clip,keepaspectratio]{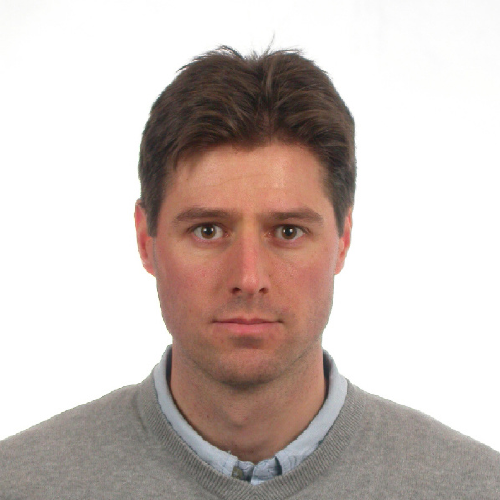}}]{Urs
Aeberhard} obtained his PhD in physics from ETH Z\"urich, Switzerland, in 2008 with a thesis on the quantum-kinetic theory of quantum-well solar cell
devices, carried out in the Condensed Matter Theory Group at Paul Scherrer Institute under the supervision of Dr. Rudolf Morf. From 2009-2012, he was a postdoctoral researcher at the Institute of Energy and Climate 
Research 5 - Photovoltaics at Forschungszentrum J\"ulich, Germany,  where he is now leading the
activities in theoretical materials science and microscopic device simulation. The focus of his
research is on the development of theory and numerical simulations for advanced nanostructure-based solar cell devices.
\end{IEEEbiography}



%\vfill


\end{document}